\documentclass[twoside,twocolumn,9pt]{article}
\usepackage{extsizes}
\usepackage[super,sort&compress,comma]{natbib} 
\usepackage[version=3]{mhchem}
\usepackage[left=1.5cm, right=1.5cm, top=1.785cm, bottom=2.0cm]{geometry}
\usepackage{balance}
\usepackage{mathptmx}
\usepackage{sectsty}
\usepackage{graphicx}
\usepackage{physics}
\graphicspath{{}{figs/}}
\usepackage{lastpage}
\usepackage[format=plain,justification=justified,singlelinecheck=false,font={stretch=1.125,small,sf},labelfont=bf,labelsep=space]{caption}
\usepackage{float}
\usepackage{fancyhdr}
\usepackage{fnpos}
\usepackage[english]{babel}
\addto{\captionsenglish}{%
  
}
\usepackage{array}
\usepackage{droidsans}
\usepackage{charter}
\usepackage[T1]{fontenc}
\usepackage[usenames,dvipsnames]{xcolor}
\usepackage{setspace}
\usepackage[compact]{titlesec}
\usepackage{hyperref}
\hypersetup{pdfstartview={FitH},pdfpagemode={UseNone},
            colorlinks,linkcolor=blue, citecolor=blue, urlcolor=blue,
            bookmarksopen=true, pdfnewwindow=true}
\usepackage[all]{hypcap}

\definecolor{cream}{RGB}{222,217,201}

\begin{document}
\nocite{rsc-control}

\pagestyle{fancy}
\thispagestyle{plain}
\fancypagestyle{plain}{
\renewcommand{\headrulewidth}{0pt}
}

\makeFNbottom
\makeatletter
\renewcommand\LARGE{\@setfontsize\LARGE{15pt}{17}}
\renewcommand\Large{\@setfontsize\Large{12pt}{14}}
\renewcommand\large{\@setfontsize\large{10pt}{12}}
\renewcommand\footnotesize{\@setfontsize\footnotesize{7pt}{10}}
\makeatother

\renewcommand{\thefootnote}{\fnsymbol{footnote}}
\renewcommand\footnoterule{\vspace*{1pt}%
\color{cream}\hrule width 3.5in height 0.4pt \color{black}\vspace*{5pt}} 
\setcounter{secnumdepth}{5}

\makeatletter 
\renewcommand\@biblabel[1]{#1}            
\renewcommand\@makefntext[1]%
{\noindent\makebox[0pt][r]{\@thefnmark\,}#1}
\makeatother 
\renewcommand{\figurename}{\small{Fig.}~}
\sectionfont{\sffamily\Large}
\subsectionfont{\normalsize}
\subsubsectionfont{\bf}
\setstretch{1.125} 
\setlength{\skip\footins}{0.8cm}
\setlength{\footnotesep}{0.25cm}
\setlength{\jot}{10pt}
\titlespacing*{\section}{0pt}{4pt}{4pt}
\titlespacing*{\subsection}{0pt}{15pt}{1pt}

\fancyfoot{}
\fancyfoot[LO,RE]{\vspace{-7.1pt}\includegraphics[height=9pt]{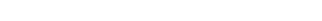}}
\fancyfoot[CO]{\vspace{-7.1pt}\hspace{13.2cm}\includegraphics{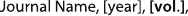}}
\fancyfoot[CE]{\vspace{-7.2pt}\hspace{-14.2cm}\includegraphics{rsc_template_details/head_foot/RF}}
\fancyfoot[RO]{\footnotesize{\sffamily{1--\pageref{LastPage} ~\textbar  \hspace{2pt}\thepage}}}
\fancyfoot[LE]{\footnotesize{\sffamily{\thepage~\textbar\hspace{3.45cm} 1--\pageref{LastPage}}}}
\fancyhead{}
\renewcommand{\headrulewidth}{0pt} 
\renewcommand{\footrulewidth}{0pt}
\setlength{\arrayrulewidth}{1pt}
\setlength{\columnsep}{6.5mm}
\setlength\bibsep{1pt}

\makeatletter 
\newlength{\figrulesep} 
\setlength{\figrulesep}{0.5\textfloatsep} 

\newcommand{\topfigrule}{\vspace*{-1pt}%
\noindent{\color{cream}\rule[-\figrulesep]{\columnwidth}{1.5pt}} }

\newcommand{\botfigrule}{\vspace*{-2pt}%
\noindent{\color{cream}\rule[\figrulesep]{\columnwidth}{1.5pt}} }

\newcommand{\dblfigrule}{\vspace*{-1pt}%
\noindent{\color{cream}\rule[-\figrulesep]{\textwidth}{1.5pt}} }

\makeatother

\twocolumn[
  \begin{@twocolumnfalse}
{\includegraphics[height=30pt]{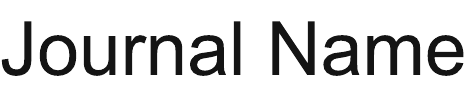}\hfill\raisebox{0pt}[0pt][0pt]{\includegraphics[height=55pt]{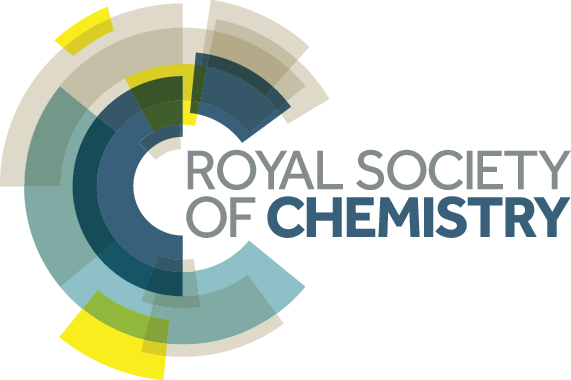}}\\[1ex]
\includegraphics[width=18.5cm]{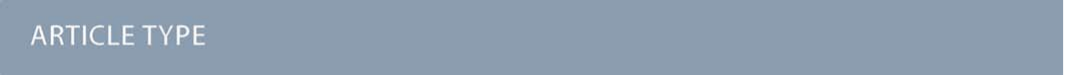}}\par
\vspace{1em}
\sffamily
\begin{tabular}{m{4.5cm} p{13.5cm} }

\includegraphics{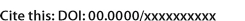} & \noindent\LARGE{\textbf{Metasurface-enabled small-satellite polarisation imaging$^\dag$}} \\
\vspace{0.3cm} & \vspace{0.3cm} \\

 & \noindent\large{Sarah E. Dean,$^{\ast}$\textit{$^{a}$} Josephine Munro,\textit{$^{a}$} Neuton Li,\textit{$^{a}$} Robert Sharp,\textit{$^{b}$}  Dragomir N. Neshev,\textit{$^{a}$} and Andrey A. Sukhorukov\textit{$^{a}$}} \\

\includegraphics{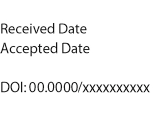} &\noindent\normalsize{
Polarisation imaging is used to distinguish objects and surface characteristics that are otherwise not visible with black-and-white or colour imaging. Full-Stokes polarisation imaging allows complex image processing like water glint filtering, which is particularly useful for remote Earth observations. The relatively low cost of small-satellites makes their use in remote sensing more accessible. However, their size and weight limitations cannot accommodate the bulky conventional optics needed for full-Stokes polarisation imaging. We present the modelling of an ultra-thin topology-optimised diffractive metasurface that encodes polarisation states in five different diffraction orders. Positioning the metasurface in a telescope's pupil plane allows the diffraction orders to be imaged onto a single detector, resulting in the capability to perform single-shot full-Stokes polarisation imaging of the Earth's surface. The five rectangular image swaths are designed to use the full width of the camera, and then each successive frame can be stitched together as the satellite moves over the Earth's surface, restoring the full field of view achievable with any chosen camera without comprising the on-ground resolution. Each set of four out of the five orders enables the reconstruction of the full polarisation state, and their simultaneous reconstructions allow for error monitoring. The lightweight design and compact footprint of the polarisation imaging optical system achievable with a metasurface is a novel approach to increase the functionality of small satellites while working within their weight and volume constraints.} \\

\end{tabular}

 \end{@twocolumnfalse} \vspace{0.6cm}

  ]

\renewcommand*\rmdefault{bch}\normalfont\upshape
\rmfamily
\section*{}
\vspace{-1cm}


\footnotetext{\textit{$^{a}$~ARC Centre of Excellence for Transformative Meta-Optical Systems (TMOS), Department of Electronic Materials Engineering, Research School of Physics, Australian National University, Canberra, ACT 2600, Australia. 
E-mail: sarah.dean@anu.edu.au}}
\footnotetext{\textit{$^{b}$~Research School of Astronomy and Astrophysics, Australian National University, Weston Creek, ACT 2611, Australia. }}

\footnotetext{\dag~Supplementary Information available: See DOI: 00.0000/00000000.}



\section{\label{sec:Introduction}Introduction}
Polarisation imaging involves measuring the orientation of the incident light's electric field vector at discrete intervals over the whole field of view of the scene. Since polarisation is a quantity of light that is independent of intensity and wavelength, polarisation imaging is a valuable tool used to distinguish objects or features that would otherwise show no contrast in black-and-white or spectral imaging. Polarisation imaging can also enhance the contrast of objects that are partially obscured by reflections, or other objects in a scene.
Since detectors are not sensitive to the polarisation state of light, multiple polarisation analyser filters need to be used to recover the incident polarisation state-- in much the same way a Bayer filter is commonly used for colour imaging.
When the full-Stokes polarisation state is imaged, including elliptical and partially polarised states, image analysis can enhance more complex features in a scene. One application of interest is satellite remote sensing of the Earth's surface, utilising the full polarisation state for applications such as analysis of chiral organic aerosols or water surface sun-glint correction \cite{Tyo2006, NASA-PACE, Zhao2016,Chowdhary2006,Wang2023}.

\begin{figure}[h]
\centering
  \includegraphics[width=\columnwidth]{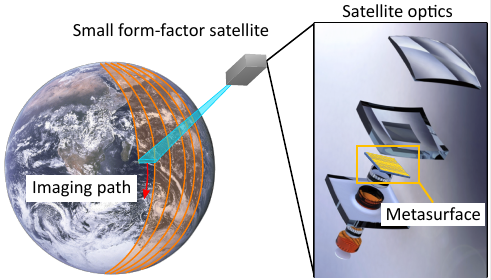}
  \caption{Illustration of the proposed satellite system. A small form-factor satellite images a strip of the Earth’s surface, forming a complete image over time. A metasurface can be placed within an existing imaging system without increasing the overall footprint. \textit{Images reproduced with permission from: NASA, image ID AS17-148-22727; Rob Sharp, CHICO hyperspectral sensor}.}
  \label{fig:conceptual}
\end{figure}

Small satellites are quickly becoming favoured for scaleable remote sensing systems, due to their accessibility and cost-effectiveness compared to traditional satellite systems \cite{Kosiak}. However, conventional full-Stokes imaging methods require either moving parts or filtering to achieve all the necessary polarisation measurements \cite{Sabatke2000, Rubin2019}, which cannot be appropriately scaled for small form-factor systems in fast-moving low-light conditions. We propose that metasurfaces can be used instead of conventional optics and placed within the same footprint of existing small-satellite systems, see Figure \ref{fig:conceptual}, decreasing the weight and volume of complex imaging systems to allow improved functionality of these satellites. Here, we present a metasurface design for polarisation imaging, adopted specifically for a small form-factor satellite.

\subsection{Metasurface optics}
Metasurfaces are nanophotonic elements consisting of a nanostructured array of metallic or dielectric scatterers on a flat substrate. They can utilise the shape and materials of the nanostructures to manipulate the phase of incident light using polarisation-sensitive resonances induced in the nanostructure \cite{Kivshar2017}.
In particular, dielectric metasurfaces have been demonstrated to efficiently behave as traditional lenses and prisms \cite{Arbabi2017,He2018}, waveplates \cite{Shi2020}, and polarisation splitters \cite{Wang2018}; with metasurfaces also combining these elements to demonstrate multifunctional capabilities. 

Single-shot polarisation imaging has been previously achieved using metasurfaces for static cameras and bright imaging targets, utilising a few different imaging setups. Metasurfaces utilising per-pixel filtering for polarisation measurements naturally create compact imaging systems \cite{Fan2023,Zuo2023}, with the metasurface extremely close to or integrated with the camera sensor. 
However, any system using filtering to form measurements inherently limits the photon flux to the sensor. While this is an effective set-up for bright imaging environments, it is crucial to maximise the transmission to the sensor in low-light conditions.
Another measurement method present in previous metasurface devices is to diffract incident light into separate measurements \cite{Arbabi2018,Rubin2019,Rubin2022,Li2023,Huang2023,Ren2023}, avoiding loss of intensity through filtering. These systems are less compact than the per-pixel filtering systems due to requiring space between the metasurface and the camera sensor; however, a diffractive metasurface-based system still provides substantial weight and size benefits over traditional optics. We choose to build on these diffractive systems, extending them to account for the additional non-trivial requirements of a remote-sensing system.

Remote sensing imposes additional challenges that aren't encountered in previous metasurface imaging works. The low-light imaging conditions ensure we want to utilise incident light as completely and efficiently as possible, and the polarisation measurements need to account for the field-of-view requirements of a moving system, with their shape and position on the sensor requiring careful design. Finally, the isolated nature of a satellite-based system necessitates that we include a method of verifying the integrity of the polarisation measurements in our design, monitoring for accumulated errors from exposure to the harsh conditions of space.

\subsection{\label{subsec:integration}Integration with other systems}

The compact form-factor of metasurface optics is well-suited to integration into existing optical systems. We choose to motivate our metasurface design by the Cubesat Hyperspectral Imager for the Coastal Ocean (CHICO) that is currently under development in Australia~\cite{matthews_demonstration_2023}. It is a small form-factor sensor (2.5U CubeSat) designed for coastal water monitoring. The CHICO system contains a primary Hyperspectral imaging channel in the visible light, and a secondary channel for temporally and spatially simultaneous glint correction of the primary channel; our metasurface will operate in this second channel.

However, the sun-glint often masks useful sub-surface ocean data. Polarisation imaging enables enhanced calibration and detection for this glint component and, therefore, allows the recording of otherwise masked subsurface ocean data.
Adding polarisation capabilities using conventional methods, such as lossy filters or bulky prisms~\cite{Sabatke2000, Chekhova2021}, would violate the transmission, size, or weight requirements of the CHICO system. Non-conventional solutions are therefore necessary for including polarisation imaging in the CHICO system, making it well-suited for demonstration of the design of a metasurface constrained by specific system requirements.

\section{Metasurface-focussed system design}
\label{sec:sys}

\subsection{\label{subsec:pushbroom}Designing for satellite field-of-view requirements}

To take advantage of the satellite movement and efficiently use sensor space, the imaging system will use pushbroom imaging. Pushbroom imaging is an established technique in remote sensing, where a narrow strip or 'swath' is imaged on a linear detector array. Multiple swaths are imaged when the satellite moves over the ground, as shown in Figure~\ref{fig:conceptual}, then stitched together to recreate the whole 2D field-of-view~\cite{NASALandSat}.  This method is often used in multispectral imaging, where a swath is separately imaged for different wavelength channels. A prominent measurement technique includes beam splitting onto separate linear detectors (division-of-amplitude). A variation of division-of-amplitude is called division-of-aperture, where each channel is re-imaged onto a single 2D detector. The simultaneous imaging of the swath in the different wavelength channels avoids sources of measurement error, such as illumination changing between measurements and movement within the scene. These imaging errors are prevalent in systems that consecutively image each wavelength channel with rotating filters (division-of-time). Division-of-focal plane imaging systems that use a colour filter for each pixel, for example, a Bayer filter for RGB, are seldom used for astronomical or satellite imaging technologies as they have a reduced resolution capability.

We have applied the principle of division-of-aperture to polarisation remote sensing, utilising a metasurface to diffract the swath into polarisation channels, and onto a single camera. This would allow one to record all polarisation measurements simultaneously and efficiently.

\subsection{\label{subsec:Polarimetry} Polarimetry measurements}

For full-Stokes polarimetry, including partially polarised states, a minimum of four polarisation measurements are required. The optimum measurements for a four-measurement full-Stokes polarimeter have been demonstrated extensively \cite{Azzam1988,Sabatke2000, Sabatke2000a,Ambirajan1995} and utilised in previous metasurface-based polarisation camera works \cite{Rubin2019,Rubin2022,Li2023}. 

The Stokes vector representation of polarisation is $\vec{S}=(S_0,\;S_1,\;S_2,\;S_3)$, where $S_0$ is the total light intensity, and $S_1$, $S_2$, and $S_3$ represent the polarisation state excluding overall phase. For an incident Stokes polarisation, $\vec{S}$ and output intensity measurements, $\vec{I}$, the relationship between these two vectors is given by:
\begin{equation}
    \label{eq:fwdpol}
    \boldsymbol{M} \vec{S} = \vec{I},
\end{equation}
where $\boldsymbol{M}$ is the instrument matrix representing the behaviour of the polarimeter system. The instrument matrix $\boldsymbol{M}$ isn't directly chosen; rather, it is calibrated by inputting known polarisation states, and using the output intensities to solve for $\boldsymbol{M}$. The resulting system, when properly optimised, has a unique output intensity vector $\vec{I}$ for each input polarisation. Reconstruction of an unknown input polarisation is then performed using the maximum likelihood method applied to a classical system \cite{James2001,Lung2024}.

\begin{figure*}[!t]
 \centering
 \includegraphics[width=\textwidth]{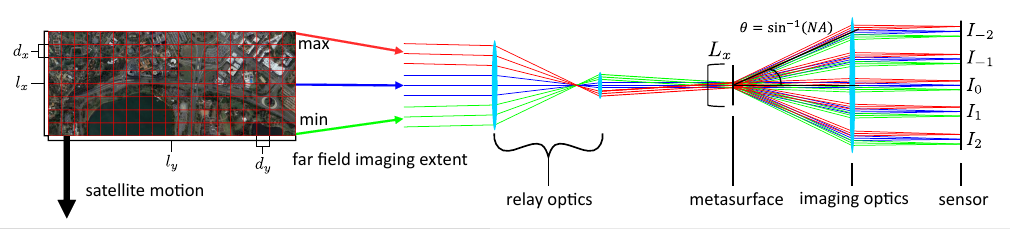}
 \caption{Illustration of the metasurface behaviour within an example of the imaging optics for the satellite The imaging target in the far-field is described by three fields within the system, corresponding to the maximum and minimum angular extent of the object, and normally incident light. Relay optics scales the incoming fields to properly incident on the metasurface. The metasurface diffracts the light into five distinct polarisation measurements, diffracting in one direction to create strips suitable for effective scanning of the Earth's surface. Finally, the target is resolved using imaging optics to form five polarisation-dependent images on the sensor.}
 \label{fig:behaviour}
\end{figure*}

We can further extend the capability of the system to include redundancy for error monitoring by using five intensity measurements instead of four. Using five measurements instead of the minimum required four measurements marginally reduces the signal-to-noise for the polarimetric measurement. The reduction in signal-to-noise ratio is considered acceptable due to the added error monitoring capability; it also increases the efficiency achievable with the metasurface.  
The loss of efficiency that would be suffered in a four diffraction order metasurface arises from zeroth-order light leakage that is typically non-negligible for actual fabricated diffractive metasurfaces, despite design optimisations that aim to eliminate zeroth-order light leakage.
It is therefore advantageous for the efficiency of our design to include the zeroth-order in optimisation as a fith measurement. 

By optimising these five measurements such that any the combination of four measurements spans the Stokes space sufficiently, we can compare the four simultaneous reconstructions to remotely monitor the metasurface polarimetry performance. For a system that is diffractively separating the intensity measurements, such as described in Section \ref{subsec:pushbroom}, the intensity measurement vector is $\vec{I}=(I_{-2},\;I_{-1},\;I_{0},\;I_{1},\;I_{2})$ and the instrument matrix $\boldsymbol{M}$ has the dimensions $5\times4$ to relate the four input Stokes parameters to the five output intensity measurements. The separate reconstructions $i$ can be represented by separate instrument matrices $\boldsymbol{M}_i$ with dimensions $4\times4$. If the reconstructions diverge in value, we can assume errors have been introduced into the system via damage or degradation of the metasurface or surrounding optics, and the imaging system requires recalibration for continued use. This is critical for a space-based system, where the performance of the imaging system can only be remotely monitored.

\subsection{\label{subsec:Res} Fourier-limited resolution}

We performed Fourier analysis on the system illustrated in Figure~\ref{fig:behaviour} to estimate the limitation of the achievable imaging resolution specifically due to the metasurface area and the periodicity of its structure, ignoring the angular imaging limitations of the system in this analysis. We assume the imaging system behaves paraxially, and effects from aberrations are ignored to further simplify and generalise the system for our estimation. We also assume the field of view in the $x$-direction is smaller than the diffraction angle. This Fourier analysis is presented in full in Section S1 of the supplementary.
Finally, we present this analysis in terms of the movement direction ($x$) values multiplied by the transverse direction ($y$) values, with the understanding that any ratio of x to y values can be applied to the final results of this analysis.

With the overall metasurface area limiting the minimum sampling size of the image, and the periodicity 
acting as a convolution kernel limiting the maximum size of the image, the total number of image samples $({l_xl_y})/({d_xd_y})$ for a metasurface diffracting to a specific numerical aperture (NA) and the overall metasurface area $L_xL_y$ is:
\begin{equation}
    \frac{L_xL_y\text{NA}^2}{4\lambda^2}>\frac{l_xl_y}{d_xd_y}.
    \label{eq:res}
\end{equation}

We consider a system with NA~=~0.29, resulting in sufficient forward diffraction orders to be captured and relayed onto the sensor without using expensive and tightly toleranced high-NA relay optics. We perform this calculation for a wavelength of $\lambda = 850$\,nm to be consistent with metasurface design choices, see Section~\ref{subsec:MSdesign} below. Emulating the CHICO imaging system, we are using an image sampling requirement of $({l_xl_y})/({d_xd_y})\geq 900 \times 100 = 90,000$ to implement pushbroom imaging as described in section \ref{subsec:pushbroom}.
We, therefore, estimate that a metasurface with area $L_xL_y> 3.1$~mm$^2$ satisfies this resolution requirement at the target metasurface periodicity and imaging wavelength, which is a feasible scale for metasurface fabrication.

\section{\label{sec:MSdesign}Metasurface design}

\subsection{\label{subsec:MSdesign}Design framework} 

There are a number of necessary considerations when selecting the imaging wavelength range for glint correction. We choose to avoid wavelengths in the visible spectrum due to the hyperspectral visible light channel of the representative CHICO system, as detailed in Section~\ref{subsec:integration}. Avoiding atmospheric absorption is also crucial for remote sensing applications, ensuring enough light can be detected. Therefore, we choose an operating bandwidth in the near-infrared range of 840-850~nm, thereby avoiding both significant atmospheric absorption \cite{Younes2021} and the wavelength range of the CHICO hyperspectral channel.

We chose a metasurface structure of 1~$\mu$m thick patterned silicon on a 460~$\mu$m sapphire substrate, with the patterned layer consisting of a 6750~$\times$~450~$\mu$m repeating design. Silicon is a frequently used material for metasurface fabrication due to its low absorption in the near-infrared, and a well-established fabrication process. 
We use a silicon thin film on a sapphire substrate due to the commercial availability of silicon-on-sapphire wafers. However, further investigation and testing are required to ensure the suitability of these materials for a space-based environment.
The metasurface periodicity produces a one-dimensional diffraction pattern in the x-direction for $\lambda = 850$~nm incident light, with second-order diffraction angles of $\theta_{\pm2}=\pm14.6^\circ$. This angle is sufficiently forward for all the diffraction orders utilised as polarisation measurement channels to be captured by traditional optics and relayed to a single camera sensor.

For the metasurface design, we utilised freeform topology optimisation to structure the silicon layer. This technique allows more complex and efficient metasurface behaviours through the development of unintuitive free-form structures. Furthermore, topology optimisation is more suited than common geometric phase approaches for our five-measurement system, as it allows the optimisation to include the zeroth order measurement without it being overexposed compared to the other orders~\cite{Sell2017,Fan2020, LalauKeraly2013,Jensen2011}. Consistent with previous works using freeform topology optimisation for metasurface design~\cite{Sell2017,Fan2020, LalauKeraly2013,Jensen2011}, we start our optimisation as a smooth, random distribution of values between the refractive index of the metasurface material and air. For a single iteration of the optimisation process, we perform both forward and adjoint simulations of the metasurface, utilising Lorentz reciprocity to numerically calculate the gradient in transmission as a function of the refractive index for every point of the metasurface simultaneously. We then use a gradient descent method to optimise the metasurface design towards a figure of merit describing the metasurface behaviour, including binarisation and robustness functions such as blurring to converge to a final, discrete and fabricable design after 300 iterations. We further performed multiple optimisation runs to find the highest performance for our final design.

We performed simulations of the metasurface behaviour using the rigorous coupled-wave analysis package RETICOLO~\cite{Hugonin2022} during the topology optimisation process. The instrument matrices $\boldsymbol{M}_i$ for each possible reconstruction $i$, as described in Section \ref{subsec:Polarimetry}, are calculated from these simulations, and the performance of these instrument matrices for polarisation reconstruction is maximised with each optimisation iteration to ensure they can be used independently for measuring polarisation. 

One established method of maximising the polarisation reconstruction performance is to minimise the condition number of the instrument matrix~\cite{Azzam1988,Ambirajan1995, Sabatke2000,Sabatke2000a, Peinado2010}:
\begin{equation}
    \text{cond}(M_i) = \frac{\text{max}_j(\mu_{ij})}{\text{min}_j(\mu_{ij})},
    \label{eq:cond}
\end{equation}
where $\mu_{ij}$ are the singular values of $\boldsymbol{M}_i$. As the maximum singular value of the instrument matrix is bounded by the incident light intensity, and maximising all singular values maximises the transmission of light through the system, we define the target figure-of-merit (FoM) as equal to 
the minimum singular value:
\begin{equation}
    \text{FoM}_i = \text{min}_{j}(\mu_{ij}).
    \label{eq:minsini}
\end{equation}

The full figure-of-merit used for maximising the polarimetry performance of all five possible reconstructions $i$ simultaneously is the product of Eq.~(\ref{eq:minsini}) for each reconstruction:
\begin{equation}
    \text{FOM}=\prod_i^5\text{min}_{j}(\mu_{ij}) .
\end{equation}


\subsection{\label{subsec:Metasurface} Optimisation results}

\begin{figure}[h]
 \centering
 \includegraphics[width=\columnwidth]{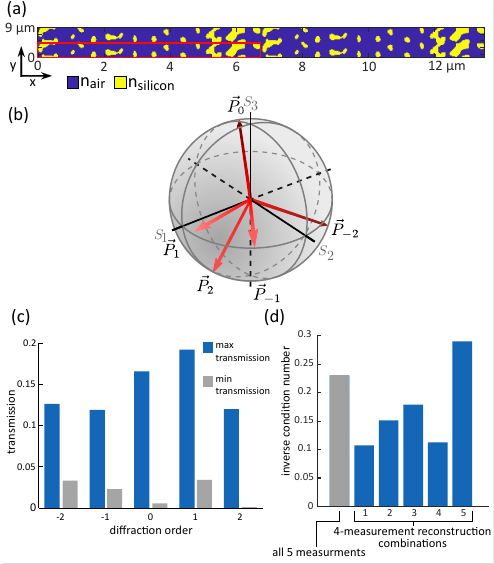}
 \caption{
 (a) Optimised metasurface design for polarisation imaging for $1~\mu$m thick patterned silicon on a $460~\mu$m sapphire substrate. A single period is 6750~nm by 450~nm, indicated in red, and repeated across the whole metasurface. 
 (b) The partial polarisers associated with each diffracted measurement can be plotted on a sphere, where the direction indicates the polarisation of the polariser analogous to Stokes vectors, and the magnitude of the vectors indicates the effectiveness of the polariser 
 (c) Transmission of the ideal polarisation state and the orthogonal state for each diffraction order gives the maximum and minimum transmission possible for the order. The contrast between the maximum and minimum transmission indicates the polarisation sensitivity of the diffraction orders. 
 (d) The inverse condition number is one measure of how robustly the polarisation measurements span the Stokes space. The inverse condition number of all possible reconstructions demonstrates individual reconstructions sufficiently span the space for 
 polarisation reconstruction.}
 \label{fig:optimsation}
\end{figure}

Following the optimisation method outlined in Section \ref{subsec:MSdesign}, our final, highest-performing metasurface design is shown in Figure~\ref{fig:optimsation}(a).

Each diffraction order $i$ may be considered an independent partial polariser with polarisation, $\vec{P}_i$ plotted as a vector on a Bloch sphere. The direction and magnitude of this vector represent the direction and degrees of polarisation, respectively. We create this representation by following the method outlined in Lung et al. \cite{Lung2024}, which is summarised here. We perform a singular value decomposition of the transfer matrix $T_m$ of each diffraction order $m$ to give
\begin{equation}
     \mathbf{T}_m = \mathbf{U}_m\mqty(\xi_{m,1}&0\\0&\xi_{m,2})\left(\mathbf{V}_m\right)^\dagger ,
     \label{eq:transfer}
\end{equation}
where $\xi_{m,1},\xi_{m,2}\geq0$ are the singular values, ordered such that $\xi_{m,1}\geq\xi_{m,2}$, and both $\mathbf{U}_m$ and $\mathbf{V}_m$ are unitary matrices. We note that the columns of $\mathbf{V}_m=\left[\Vec{V}_{m,1},\Vec{V}_{m,2}\right]$ form the basis states of the partial polariser. Converting them to a Stokes basis allows us to visualise these states on a Stokes-like sphere. The magnitude of the vector is given by:
\begin{equation}
    R=1-\frac{\xi_{m,2}^2}{\xi_{m,1}^2},
    \label{eq:magnitude}
\end{equation}
where $R=1$ represents a fully polarised output, and $R<1$ represents a partial polariser with finite extinction ratio $(1-R)^{-1}$. This representation of the partial polarisers is shown in Figure~\ref{fig:optimsation}(b) for all diffraction orders, demonstrating that the measurements from the five diffraction orders span the Stokes space. This procedure ensures the incident light is properly sampled for performing polarimetry.

The selectivity of the diffraction orders to different polarisations is quantised by the transmission of the ideal polarisation state $\vec{P}_i$, from Figure~\ref{fig:optimsation}(b), and its orthogonal state for each diffraction order. The fraction of light transmitted for these states is given in Figure~\ref{fig:optimsation}(c), demonstrating sufficient selectivity to polarisation to distinguish the input states. It should be noted that the minimum transmissions are expected to be non-zero; as the system is overdetermined, the measurements performed by the metasurface cannot form an orthogonal basis.

As noted in Section~\ref{subsec:MSdesign}, the inverse condition number is a frequently used measure of how robust the measured polarisation states span the Stokes space. Figure~\ref{fig:optimsation}(d) plots the inverse condition number for every possible combination of 4 out of 5 polarisation measurements, and the inverse condition number using all 5 measurements. These results indicate accurate polarisation reconstruction is achievable for each set of polarisation measurements, enabling the comparison of reconstructions for error monitoring. The polarimetry performance is maintained over a wavelength range of $(842-852)$~nm, and for a field-of-view of $8^\circ$ by $20^\circ$, ensuring the metasurface performance is robust to external factors and can be used for imaging applications. Full details on the metasurface angular and wavelength performance can be found in the supplementary (Section S3).

\begin{figure*}[!t]
 \centering
 \includegraphics[width=0.99\textwidth]{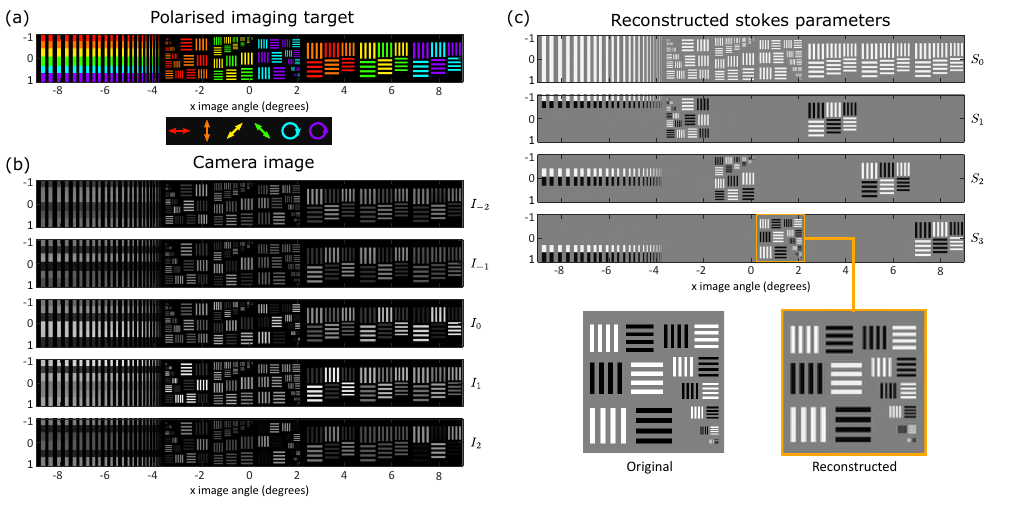}
 \caption{Simulation of imaging and reconstructing a polarised scene, in terms of the image angle at the metasurface. The equivalent angle in the far-field is dependent on the scaling factor of the optics before the metasurface. The metasurface is simulated as a 5.28~mm $\times 0.587$~mm aperture. 
 (a) The polarised imaging target is colour-coded by the input polarisations. 
 (b) The diffracted intensity measurements, as seen by the camera sensor, with the variation in intensities demonstrating polarisation-dependent metasurface behaviour over a range of incident angles. 
 (c) Reconstruction of the Stokes polarisation state, demonstrating the Fourier limited resolution for the simulated metasurface. The reconstruction resolves points with a $0.023^\circ$ separation incident on the metasurface.}
 \label{fig:imaging}
\end{figure*}

\section{\label{sec:SimIm}Polarisation imaging and reconstruction} 

Next, we verified the imaging and error monitoring performance of the metasurface through simulations with known input polarisations. We ensure that we satisfy satellite-based design requirements, such as field-of-view, resolution, and long-term system robustness.
To simulate the imaging behaviour of the metasurface, we created a digital imaging target, Figure \ref{fig:imaging}(a), for testing the polarisation reconstruction and spatial resolution performance of the metasurface over a realistic field-of-view for pushbroom satellite imaging. The target consists of differently polarised areas that span the Stokes space to test the full polarisation response, and features of different sizes and separations to test the spatial resolution of the metasurface. The target corresponds to a field-of-view of $\pm9^\circ$ by $\pm1^\circ$ at the metasurface, an angular range that preserves the overall metasurface reconstruction behaviour without overlapping measurements on the camera sensor, while also matching the imaging aspect ratio required by the CHICO sensor, outlined in Section~\ref{subsec:Res}.


We demonstrate the metasurface imaging performance in terms of the angle of incidence on the metasurface, as it allows our analysis to be performed independently of the surrounding imaging and relay optics in the complete system, thereby creating a generalised result for future implementation. We also ignore any possible effects from lens aberrations in the system to maintain this generalised result; a full analysis of the effect of the surrounding optics on the metasurface performance will be performed alongside laboratory characterisation and satellite system prototyping.

The dimensions of the simulated metasurface are $5.28\times 0.587$~mm$ = 3.1$~mm$^2$, such that the metasurface has a 9:1 aspect ratio to address the field-of-view. The metasurface area equals the minimum area calculated in Section \ref{subsec:Res} for the required imaging resolution using the CHICO project guidelines.


There are two components to simulating the imaging performance of the metasurface: the effect of the finite-sized metasurface as an aperture on the image and the polarisation-dependent diffraction of the incident image into separate imaging channels. These operations are commutative as they independently address different components of the input image.
The precise polarisation response of the metasurface is dependent on the angle of incidence of the input light. We calculated the metasurface polarisation behaviour under various incident angles spanning the target field-of-view. We used this polarisation behaviour to calculate the polarisation-dependent transmission to each diffraction order for every point of the input. The resulting simulated camera image of the output diffraction orders is shown in Figure~\ref{fig:imaging}(b), with each diffraction order having a unique intensity distribution based on its selectivity to polarisation, described in Section~\ref{subsec:Metasurface}. The transmission intensity of any singular polarisation does vary over the camera image due to the angular dependence of the metasurface; however, this can be easily accounted for by including this angularly dependent behaviour during the polarisation reconstruction process. These camera measurements are successfully reconstructed to give the original polarisation in terms of the Stokes parameters, see Figure~\ref{fig:imaging}(c).

The metasurface is in the pupil plane of any surrounding optics. Therefore, we can use Fourier analysis to calculate and analyse the effect of the metasurface as an aperture on the imaging resolution. We first calculate the Fourier transform of the far-field input image, then take the product of the pupil plane light field with the aperture, and perform an inverse Fourier transform to get the final simulated camera image of the input. The effect of the aperture is present in the camera image in Figure~\ref{fig:imaging}(b), and is highlighted in the reconstructed Stokes parameters in Figure~\ref{fig:imaging}(c). The reconstruction achieves an angular resolution of approximately $70.2$~arcsec separation between distinguishable points, in agreement with the Fourier-limited resolution calculation in Section~\ref{subsec:Res}.

\begin{figure}[h]
\centering
  \includegraphics[width=\columnwidth]{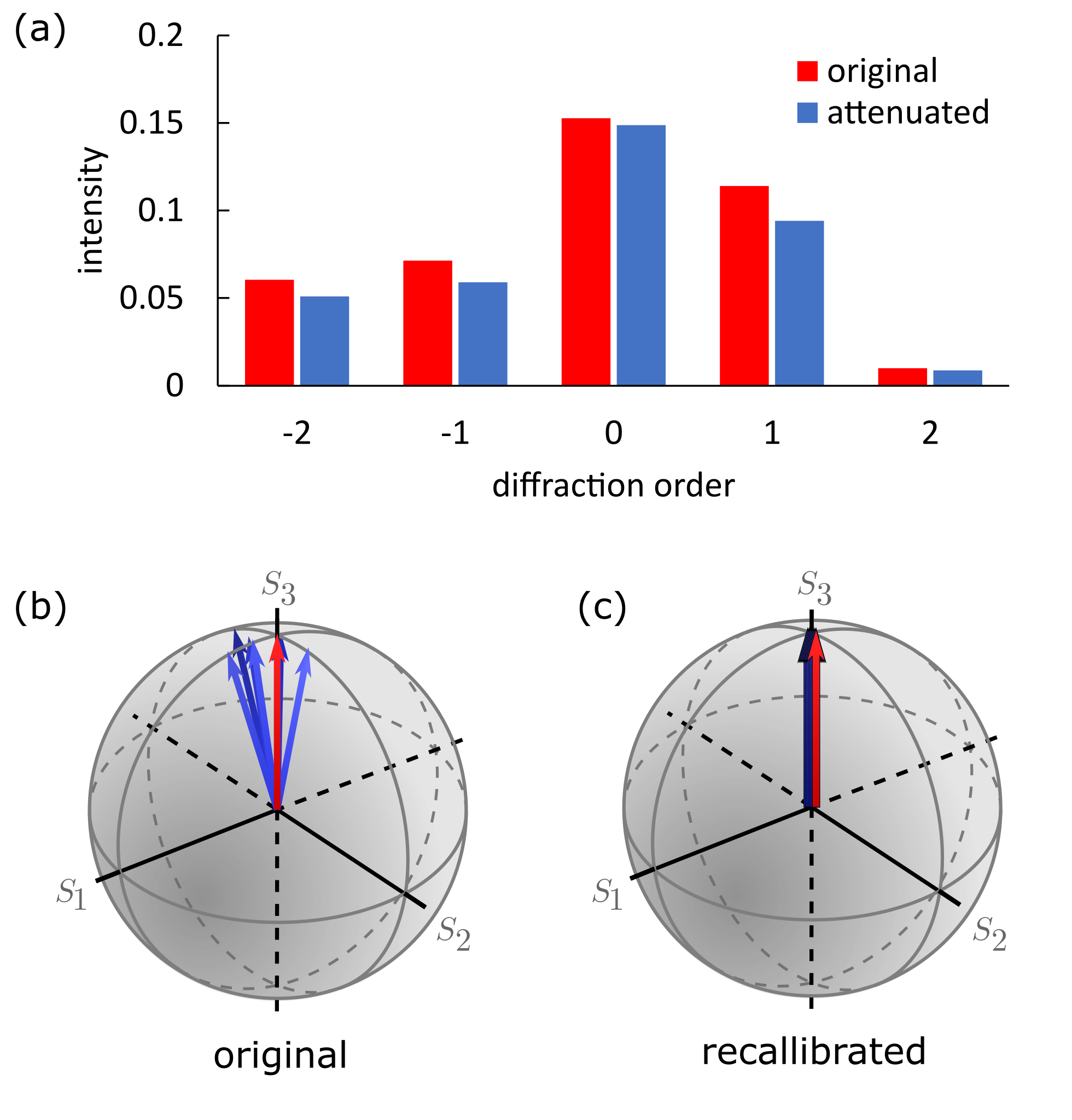}
  \caption{
  (a) Measurements recorded by the camera for incident light with a polarisation of $(1,0,0,1)$ for an imaging system working precisely as designed (red), and for a system with errors introduced (blue), where we apply nonpolarising attenuation of a random value up to 10\% to each measurement independently. 
  (b) The original incident state (red) and the reconstructions of the poisoned states using every possible combination of four or five measurements. The reconstructed vectors diverge in different directions from the incident state, indicating errors are present. 
  (c) Recalibration of the reconstruction matrices allows us to correctly reconstruct the polarisation with the same imaging system}
  \label{fig:poisoned}
\end{figure}

Finally, we demonstrated the performance of the metasurface in a system with errors introduced into our measurements. The errors represent effects such as degradation of the metasurface or minor damage to the imaging optics. For a simplified example, we apply nonpolarising attenuation of a random value up to 10\% to each measurement independently, and record the change in the measurement for the same incident polarisation. The results of these simulations are shown in Figure~\ref{fig:poisoned}(a). We used each subset of 4 measurements to perform different reconstructions, utilising the maximum likellihood method to ensure the reconstructed values are physical~\cite{James2001, Lung2024}. Comparing these measurements shows that errors in the system result in reconstructions with diverging polarisations, indicating the presence of errors, as seen in Figure \ref{fig:poisoned}(b). Re-characterising the metasurface using input light with known polarisations and recalculating the instrument matrices for each possible polarisation reconstruction, as described in Section~\ref{subsec:Polarimetry}, allows the entire system to be re-calibrated remotely, see Figure~\ref{fig:poisoned}(c). This ability to account for degradation of the system, including errors arising from degradation of components other than the metasurface, and lack of reliance on the metasurface structure precisely matching the designed pattern, ensures the metasurface is robust to long-term use in a satellite-based system.


\section{Conclusions}


We have presented a topology-optimised design for a metasurface that satisfies the stringent size, weight, and power constraints of optical polarimetry systems for small-satellite systems. Importantly this design allows for single-shot, full-Stokes polarisation imaging with error monitoring in a very compact optical footprint. Our metasurface polarimeter is deliberately overdetermined: the five diffraction orders are polarisation encoded to allow for four simultaneous polarisation reconstructions,  enabling long-term error monitoring and recalibration of the system behaviour. Our simulation results show the polarimetric resolution performance of the metasurface is maintained over $\pm$10$^\circ$ field-of-view, meaning a satellite system can have an adequate swath width for imaging. 

The amount of optical components and complexity that is conventionally required for full-Stokes polarisation imaging confines it to only large satellites, however, our metasurface design and computational results show a full-Stokes polarisation imaging system is viable for much more commercially available small-satellite systems. Having full-Stokes polarisation imaging capability is particularly important for remote Earth imaging small-satellites, where both a compact form factor and imaging capability are essential for mission success. 

The metasurface approach to enhancing small-satellite Earth observation also has the potential to enable complex processing such as water glint removal and edge detection for future small-satellite missions.




\section*{Author contributions}

Sarah Dean contributed to methodology, optimisation, analysis, and manuscript writing. Josephine Munro contributed to system design and manuscript reviewing. Neuton Li contributed to methodology and manuscript reviewing. Robert Sharp, Dragomir Neshev and Andrey Sukhorukov contributed to research conceptualisation, supervision, and manuscript reviewing.


\section*{Conflicts of interest}

There are no conflicts to declare.

\section*{Data availability}

This study was carried out using the publicly available software package RETICOLO by J.-P.~Hugonin and P.~Lalanne\cite{Hugonin2022}, found at \url{https://doi.org/10.5281/zenodo.5905381}. The version of the code employed for this study is version V9.

The data generated and analysed in this study is included as part of the main article and as part of the Supplementary Information.

\section*{Acknowledgements}
This work was supported by the Australian Research Council under the Centre of Excellence for Transformative Meta-Optical Systems (TMOS) (CE200100010) and a research grant (NI210100072).



\balance


\bibliography{satellite} 
\bibliographystyle{rsc} 
\end{document}